\documentclass{PoS}
\usepackage{axodraw2}
\usepackage{amsmath}

\title{Analytical results for hadronic contributions to the muon $g-2$}

\ShortTitle{Analytical results for $(g-2)_\mu$}

\author{\speaker{Johan Bijnens}\\
        Lund University\\
        E-mail: \email{bijnens@thep.lu.se}}

\author{Nils Hermansson-Truedsson\thanks{Present address: Universit\"at Bern, E-mail: nils@itp.unibe.ch}\\
        Lund University\\
        E-mail: \email{nils.hermansson-truedsson@thep.lu.se}}

\author{Antonio Rodr\'\i guez-S\'anchez\\
        Lund University\\
        E-mail: \email{antonio.rodriguez@thep.lu.se}}

\abstract{This talk discusses two analytical calculations relevant for the
Standard Model calculation of the muon $g-2$. The first part is the recent
derivation of the quark-loop as the first term in a well-defined
operator-product expansion for the short-distance part of the hadronic
light-by-light contribution, as well as the calculation of the next term.
The second part is the calculation of finite volume effects relevant for
lattice QCD calculations of the electromagnetic contribution to the
lowest-order hadronic vacuum-polarization contribution and the proof they only
start at $1/L^3$.}

\FullConference{
European Physical Society Conference on High Energy Physics - EPS-HEP2019 -\\
			10-17 July, 2019\\
			Ghent, Belgium}

\renewcommand{\logo}
{\raisebox{-5mm}
{\parbox{\textwidth}{\begin{flushright}
LU TP 19-49\\
October 2019\\
\end{flushright}
\includegraphics{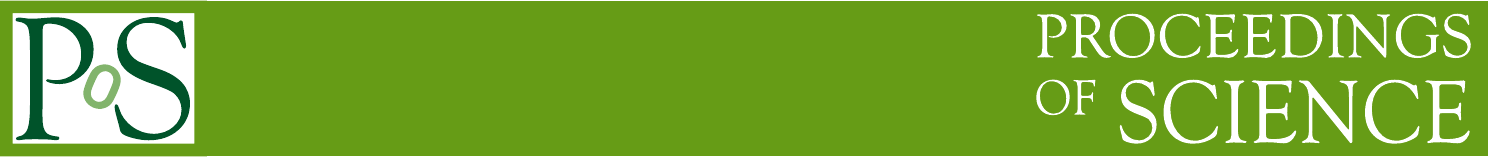}}
}}
\FullConference{{\rm To be published in the proceedings of:\\}
European Physical Society Conference on High Energy Physics - EPS-HEP2019 -\\
			10-17 July, 2019\\
			Ghent, Belgium}

\begin{document}

\section{Introduction}

The muon anomalous magnetic moment is among the most precisely measured
quantities in particle physics. The final result~\cite{Bennett:2006fi} differs
from the Standard Model prediction by 3.5-4~$\sigma$,
see e.g.~\cite{Jegerlehner:2017lbd}. An ongoing experiment at FNAL
aims to increase the measurement precision by about a factor of four and
there is an experiment under development
with an innovative approach at J-PARC.

\begin{figure}[t]
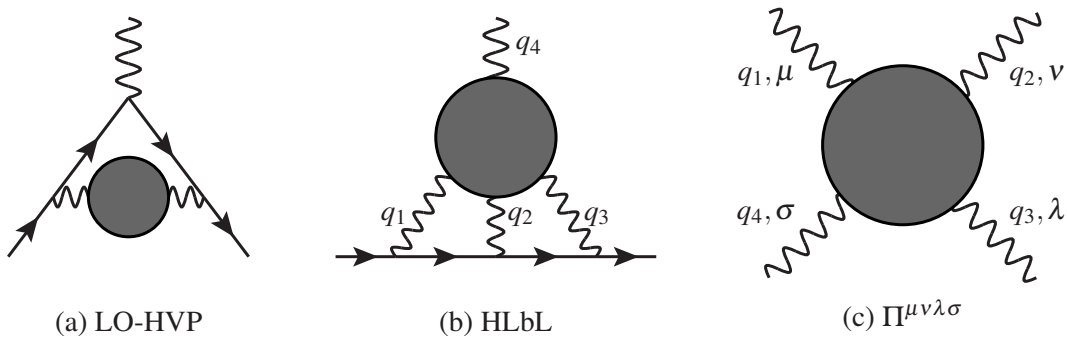

\begin{center}
\begin{axopicture}(60,90)(0,-20)
\SetScale{1.5}
\SetWidth{0.75}
\Photon(30,60)(30,40){3}{3.5}
\ArrowLine(0,0)(11.25,15)
\ArrowLine(11.25,15)(30,40)
\ArrowLine(30,40)(48.75,15)
\ArrowLine(48.75,15)(60,0)
\Photon(11.25,15)(20,15){-2.5}{2.}
\Photon(40,15)(48.75,15){2.5}{2.}
\GCirc(30,15){10}{0.4}
\Text(30,-20)[b]{(a) LO-HVP}
\end{axopicture}
\hspace*{2cm}
\setlength{\unitlength}{1pt}
\begin{axopicture}(80,90)(0,-20)
\SetScale{1.5}
\SetWidth{0.75}
\Photon(40,60)(40,40){3}{3.5}
\ArrowLine(0,0)(15,0)
\ArrowLine(15,0)(40,0)
\ArrowLine(40,0)(65,0)
\ArrowLine(65,0)(80,0)
\Photon(15,0)(40,40){2}{9}
\Photon(40,0)(40,40){-2}{8}
\Photon(65,0)(40,40){-2}{9}
\GCirc(40,30){15}{0.4}
\Text(18,10)[r]{$q_1$}
\Text(43,10)[l]{$q_2$}
\Text(62,10)[l]{$q_3$}
\Text(45,53)[l]{$q_4$}
\Text(40,-20)[b]{(b) HLbL}
\end{axopicture}
\hspace*{2.7cm}
\setlength{\unitlength}{1pt}
\begin{axopicture}(100,120)(0,-12)
\SetScale{1.0}
\SetWidth{1.125}
\Photon(0,0)(100,100){4}{15}
\Photon(0,100)(100,0){4}{15}
\GCirc(50,50){30}{0.4}
\Text(90,20)[lb]{$q_3,\lambda$}
\Text(90,80)[lt]{$q_2,\nu$}
\Text(10,80)[rt]{$q_1,\mu$}
\Text(10,20)[rb]{$q_4,\sigma$}
\Text(50,-20)[b]{(c) $\Pi^{\mu\nu\lambda\sigma}$}
\end{axopicture}

\end{center}
\caption{\label{fig:hadronic} Hadronic contributions to the muon $g-2$.(a) HVP, (b) HLbL, (c) The vector four-point function}
\end{figure}

It is thus imperative that the theoretical error should be brought down to the
same level of precision. The theoretical error is dominated by the
hadronic contributions depicted schematically in Figs.~\ref{fig:hadronic}a
and b. 

The hadronic light-by-light (HLbL) contribution and the recent progress
on its short-distance part~\cite{Bijnens:2019ghy} is discussed in
Sect.~\ref{sec:HLbL}. The main conclusion is that the quark-loop as often used
in this respect really is the first term in a proper short-distance expansion.
 The lowest-order hadronic vacuum-polarization (HVP)
and our estimate of electromagnetic finite volume contributions~\cite{Bijnens:2019ejw}
is discussed in Sect.~\ref{sec:HVP}. The main conclusion here is that these
corrections start at order $1/L^3$ and not at order $1/L^2$. This is
a general property and we check it explicitly in the case of scalar QED.

\section{Hadronic light-by-light}
\label{sec:HLbL}

This is the contribution depicted in Fig.~\ref{fig:hadronic}b. The problem
here is this contribution mixes low- and high-energy contributions and thus
as a consequence double-counting between hadron-exchanges versus quark-gluon
parts is an important aspect to consider. The hadronic object needed is the vector
four-point function depicted in Fig.~\ref{fig:hadronic}c. In general this
has 138 Lorentz-structures but in four dimensions there are 41 independent
combinations. Of these 12 combinations are needed in the limit $q_4\to 0$.
A full analysis and references to earlier work can be found
in~\cite{Colangelo:2017fiz}. Based on the dispersive method 
of~\cite{Colangelo:2017fiz} the long-distance
contribution can clearly be brought under control. The major remaining part is
now the intermediate and short-distance behaviour. The quark-loop
has been used in this context before~\cite{Bijnens:1995xf,Bijnens:2007pz}.
The operator product expansion (OPE) has been used in the HLbL context in many ways
before, mainly to put restrictions on the behaviour of form-factors of hadrons
but also to put constraints on the full vector four-point function
in the kinematic limit $Q_1^2\approx Q_2^2 \gg Q_3^2$ with $q_i^2=-Q_i^2$
and all momenta Euclidean~\cite{Melnikov:2003xd}.

\begin{figure}[t]
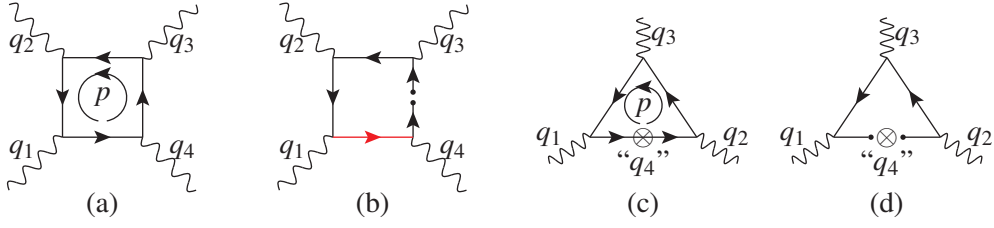

\begin{center}
\begin{axopicture}(70,80)(0,-10)
\Photon(0,0)(20,20){2.5}{3.5}
\Photon(0,70)(20,50){2.5}{3.5}
\Photon(70,70)(50,50){2.5}{3.5}
\Photon(70,0)(50,20){2.5}{3.5}
\ArrowLine(20,20)(50,20)
\ArrowLine(50,20)(50,50)
\ArrowLine(50,50)(20,50)
\ArrowLine(20,50)(20,20)
\Text(60,15)[l]{$q_4$}
\Text(10,15)[r]{$q_1$}
\Text(10,55)[r]{$q_2$}
\Text(60,55)[l]{$q_3$}
\Arc[arrow](35,35)(9,-80,260)
\Text(35,35){$p$}
\Text(35,-10)[b]{(a)}
\end{axopicture}
\hspace*{1cm}
\begin{axopicture}(70,80)(0,-10)
\Photon(0,0)(20,20){2.5}{3.5}
\Photon(0,70)(20,50){2.5}{3.5}
\Photon(70,70)(50,50){2.5}{3.5}
\Photon(70,0)(50,20){2.5}{3.5}
\ArrowLine(50,20)(50,33)
\ArrowLine(50,37)(50,50)
\Vertex(50,33){1}
\Vertex(50,37){1}
\ArrowLine(50,50)(20,50)
\ArrowLine(20,50)(20,20)
\Text(60,15)[l]{$q_4$}
\Text(10,15)[r]{$q_1$}
\Text(10,55)[r]{$q_2$}
\Text(60,55)[l]{$q_3$}
{\SetColor{Red}\ArrowLine(20,20)(50,20)}
\Text(35,-10)[b]{(b)}
\end{axopicture}
\hspace*{1cm}
\begin{axopicture}(60,80)(0,-20)
\Photon(0,0)(15,10){2.5}{3.5}
\ArrowLine(15,10)(35,10)
\ArrowLine(35,10)(55,10)
\Text(35,10){$\otimes$}
\Text(35,5)[t]{``$q_4$''}
\ArrowLine(55,10)(35,40)
\ArrowLine(35,40)(15,10)
\Photon(55,10)(70,0){2.5}{3.5}
\Photon(35,40)(35,55){2.5}{3.5}
\Text(5,10)[r]{$q_1$}
\Text(65,10)[l]{$q_2$}
\Text(38,48)[l]{$q_3$}
\Arc[arrow](35,22)(7,-80,260)
\Text(35,21){$p$}
\Text(35,-20)[b]{(c)}
\end{axopicture}
\hspace*{1cm}
\begin{axopicture}(60,80)(0,-20)
\Photon(0,0)(15,10){2.5}{3.5}
\Line(15,10)(28,10)
\Vertex(29,10){1}
\Line(41,10)(55,10)
\Vertex(41,10){1}
\Text(35,10){$\otimes$}
\Text(35,5)[t]{``$q_4$''}
\ArrowLine(55,10)(35,40)
\ArrowLine(35,40)(15,10)
\Photon(55,10)(70,0){2.5}{3.5}
\Photon(35,40)(35,55){2.5}{3.5}
\Text(5,10)[r]{$q_1$}
\Text(65,10)[l]{$q_2$}
\Text(38,48)[l]{$q_3$}
\Text(35,-20)[b]{(d)}
\end{axopicture}
\end{center}
\caption{\label{fig:quarkloop} The quark-loop contribution: (a) the standard
one, (b) a divergent OPE correction, (c) the quark-loop in the OPE in a background field, (d) the next order in the background field OPE with an insertion of the induced condensate.}
\end{figure}

We look now at the case where $Q_1^2,Q_2^2,Q_3^2\gg \Lambda_{QCD}^2$.
The quark-loop contribution depicted in Fig.~\ref{fig:quarkloop}a is finite in
this limit but if we try to calculate higher order terms in the OPE,
e.g. the $m_q\langle\bar q q\rangle$ contribution we get divergences when
putting $q_4\to 0$. This is clearly seen in Fig.~\ref{fig:quarkloop}b where
the red propagator diverges in this limit.
A similar problem occurred in QCD sum rules for the baryon magnetic
moments~\cite{Ioffe:1983ju} and we have adapted their method to the case at
hand. We use an OPE in the presence of the background field and use the
radial gauge for the background field. The latter allows to immediately take
the limit $q_4\to0$ and the lowest order contribution is depicted
in Fig.~\ref{fig:quarkloop}c. The propagator with the crossed circle is the
quark propagator in the background field. This contribution is exactly the
same as the usual quark-loop calculated via
Fig.~\ref{fig:quarkloop}a~\cite{Bijnens:2019ghy}. The next term is
proportional to the induced condensate
$\langle \bar q\sigma_{\alpha\beta} q\rangle \equiv  e_q F_{\alpha\beta} X_q$.
The quantities $X_q$ have been determined in lattice QCD and are about
40~MeV~\cite{Bali:2012zg}. The contribution to HLbL can be written using six
functions~\cite{Colangelo:2017fiz} and for the next term these are given by\footnote{These differ by a factor of $-2$ from the preliminary results shown at the conference.}
\begin{align}
\hat\Pi_1=&\,m_q X_q e_q^4\frac{-4(Q_1^2+Q_2^2-Q_3^2)}{Q_1^2 Q_2^2 Q_3^4},&
\hat\Pi_7=&\,0,\\
\hat\Pi_4=&\,m_q X_qe_q^4\frac{8}{Q_1^2 Q_2^2 Q_3^2},&
\hat\Pi_{17}=&\,m_q X_q e_q^4\frac{8}{Q_1^2 Q_2^2 Q_3^4},\\
\hat\Pi_{54}=&\,m_q X_qe_q^4\frac{-4(Q_1^2-Q_2^2)}{Q_1^4 Q_2^4 Q_3^2},&
\hat\Pi_{39}=&\,0.
\end{align}
Note that they are suppressed w.r.t. the pure quark-loop by two powers of the
hard scales, not four as would be expected from a first contribution
arising from $\langle\alpha_S G^2\rangle$ or $m_q\langle\bar qq\rangle$.

Numerical results for the quark-loop and the next term in the OPE for 
$Q_1,Q_2,Q_3 \ge Q_\text{min}$,
$m_u=m_d=m_s=0$ for the quark-loop and
$m_u=m_d=5$~MeV and $m_s=100$~MeV for $m_q X_q$ are given in Tab.~\ref{tab:HLbL}.
\begin{table}
\begin{center}
\begin{tabular}{cccc}
\hline
$Q_\text{min}$ & quark-loop & $m_u X_u+m_d X_d$ & $m_s X_s$\\
\hline
\rule{0cm}{2.5ex} 1~GeV & $17.3\times 10^{-11}$ & $5.40\times 10^{-13}$ & $8.29\times 10^{-13}$\\
\rule{0cm}{2.5ex} 2~GeV & $4.35\times 10^{-11}$ & $3.40\times 10^{-14}$ & $5.22\times 10^{-14}$\\
\hline
\end{tabular}
\end{center}
\caption{\label{tab:HLbL} Numerical results for the quark-loop and the next term
in the OPE.}
\end{table}
To be noted that for the quark-loop the contribution
above 1~GeV is still 15\% of the total value of HLbL estimated.
The quark-loop goes as $1/Q_\text{min}^2$ and the $m_q X_q$ contribution goes
as $1/Q_\text{min}^4$. This can be shown using dimensional arguments.
The second term is very small since both the values for $X_q$ and the relevant
quark-masses are very small. Higher order terms will not have this suppression
and are under investigation.

\section{Hadronic vacuum-polarization}
\label{sec:HVP}

The lowest-order hadronic-vacuum-polarization contribution is the largest
hadronic contribution to the muon anomalous magnetic moment. It can be
determined via a dispersion relation directly from experiment and the precision
of these determinations is about 0.5\%. Low-energy QCD can also be studied
via lattice QCD. The present accuracy is a few \%,
see e.g.~\cite{Meyer:2018til}, but the precision is expected to improve in the
future. The precision needed requires that electromagnetic and other isospin
breaking corrections are taken into account. The finite volume
corrections are known to two-loop order in Chiral Perturbation
Theory~\cite{Bijnens:2017esv} but the electromagnetic finite volume corrections
can be much larger since they are only suppressed by powers of the lattice size
$L$, $1/L^n$ rather than exponentially, $\exp(-mL)$.

The main object here is the vector two-point function
\begin{align}
\Pi^{\mu\nu}_{EM}(q) = i\int d^4x e^{i q\cdot x}\langle0\vert T(j^\mu_a(x)j^{\nu\dagger}_b(0)
\vert 0\rangle
\end{align}
with $j^\mu_{EM} = ({2}/{3}) j^\mu_U-({1}/{3}) j^\mu_D-({1}/{3}) j^\mu_S$ and
$j^\mu_Q = \overline q\gamma^\mu q$. In the continuum we can write
$\Pi^{\mu\nu}_{EM}(q) = \left(q^\mu q^\nu-q^2 g^{\mu\nu}\right) \Pi_{EM}(q^2)$
and the contribution to the anomalous magnetic moment is
with a known positive weight function $v$ and $Q^2=-q^2$:
\begin{align}
a_\mu = \int_0^\infty dQ^2 v(Q^2)
\left(-\Pi(Q^2)+\Pi(0)\right)\,.
\end{align}

We first calculate the finite volume electromagnetic correction in scalar QED
with Lagrangian
\begin{align}
\mathcal{L} = \left(\partial_\mu\Phi^*+i e A_\mu\Phi^*\right)
\left(\partial_\mu\Phi-i e A_\mu\Phi\right)-m_0^2\Phi^*\Phi-\frac{1}{4}F_{\mu\nu}F^{\mu\nu}\,.
\end{align}
The $\lambda\left(\Phi^*\Phi\right)^2$ is not needed to the order we are
working. The photon corrections are calculated using the methods
of~\cite{Davoudi:2018qpl} extended to two-loop order. The integral over
photon momenta in loop integrals is replaced by the sum
\begin{align}
\int d^dk \frac{1}{k^2}\to \int dk^0 \sum_{\vec k}
\frac{1}{(k^0)^2-\vec k^2}
\end{align}
and the sum needs regularizing. In $QED_L$ we do this by dropping the
parts in the sum with $\vec k=0$. As an example take the two-loop integral
\begin{align}
S = \frac{1}{i^2}\int\frac{d^dl}{(2\pi)^d}\frac{d^dk}{(2\pi)^d}
\frac{1}{k^2(l^2-m^2)\left((k+l-p)^2-m^2\right)}\,.
\end{align}
The $l^0,k^0$ integrals are done via contour integration and we now write
$\vec k = \frac{2\pi}{L} \vec n$ and expand in $1/L$.
The $\vec k$ part can be written as
\begin{align}
\frac{1}{L^{d-1}}\sum_{\vec n\ne \vec 0}
= \int \frac{d^{d-1}k}{(2\pi)^{d-1}} +\left[\frac{1}{L^{d-1}}\sum_{\vec n\ne \vec 0}-\int \frac{d^{d-1}k}{(2\pi)^{d-1}}\right]
\end{align}
In the first term we resum the series in $1/L$ which gives the infinite volume
contribution. The quantity in brackets we define to be
$\left(1/L^{d-1}\right)\Delta^\prime_{\vec n}$ and the $1/L^n$ dependence
can be characterized via the coefficients
$c_m=\Delta_{\vec n}^\prime \frac{1}{\vert \vec n\vert^m}$. These are known numerically.

The correction to the mass is
\begin{align}
\frac{\Delta_V m^2}{m^2} = e^2 \left(\frac{4 c_2}{16\pi^2 mL}+\frac{2c_1}{16\pi^2 m^2L^2}
+\mathcal{O}\left(\frac{1}{L^4},e^{-mL}\right)\right)
\end{align}
which agrees with the known result. Numerically this can be a very large
correction. It starts at $1/L$ since the photon and the pion can both be
(almost) on-shell.

\begin{figure}[t]
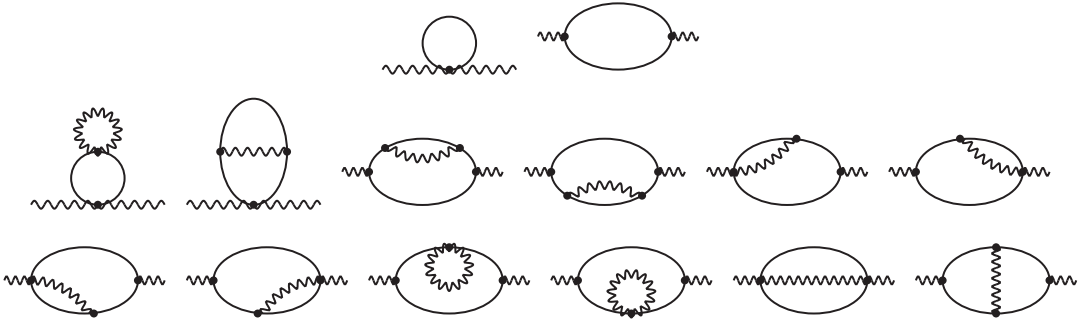

\begin{center}
\SetScale{0.5}
\begin{axopicture}(100,100)
\SetWidth{1.5}
\Photon(0,20)(50,20){3}{5}
\Photon(50,20)(100,20){-3}{5}
\Vertex(50,20){3}
\Oval(50,40)(20,20)(0)
\end{axopicture}
~~
\begin{axopicture}(120,80)(0,15)
\SetWidth{1.5}
\Photon(0,60)(20,60){3}{3}
\Oval(60,60)(25,40)(0)
\Photon(100,60)(120,60){3}{3}
\Vertex(20,60){3}
\Vertex(100,60){3}
\end{axopicture}
\\
\SetScale{0.5}
\begin{axopicture}(100,100)
\SetWidth{1.5}
\Photon(0,20)(50,20){3}{5}
\Photon(50,20)(100,20){-3}{5}
\Vertex(50,20){3}
\Oval(50,40)(20,20)(0)
\PhotonArc(50,75)(15,0,360){3}{15}
\Vertex(50,60){3}
\end{axopicture}
~~
\begin{axopicture}(100,100)
\SetWidth{1.5}
\Photon(0,20)(50,20){3}{5}
\Photon(50,20)(100,20){-3}{5}
\Vertex(50,20){3}
\Oval(50,60)(40,25)(0)
\Photon(25,60)(75,60){3}{5.5}
\Vertex(25,60){3}
\Vertex(75,60){3}
\end{axopicture}
~~
\begin{axopicture}(120,80)(0,15)
\SetWidth{1.5}
\Photon(0,60)(20,60){3}{3}
\Oval(60,60)(25,40)(0)
\Photon(100,60)(120,60){3}{3}
\Vertex(20,60){3}
\Vertex(100,60){3}
\PhotonArc(60,110)(40,230,310){3}{7.5}
\Vertex(32,78){3}
\Vertex(88,78){3}
\end{axopicture}
~~
\begin{axopicture}(120,80)(0,15)
\SetWidth{1.5}
\Photon(0,60)(20,60){3}{3}
\Oval(60,60)(25,40)(0)
\Photon(100,60)(120,60){3}{3}
\Vertex(20,60){3}
\Vertex(100,60){3}
\PhotonArc(60,10)(40,50,130){3}{7.5}
\Vertex(32,42){3}
\Vertex(88,42){3}
\end{axopicture}
~~
\begin{axopicture}(120,80)(0,15)
\SetWidth{1.5}
\Photon(0,60)(20,60){3}{3}
\Oval(60,60)(25,40)(0)
\Photon(100,60)(120,60){3}{3}
\Vertex(20,60){3}
\Vertex(100,60){3}
\PhotonArc(20,110)(50,270,330){3}{7.5}
\Vertex(67,85){3}
\end{axopicture}
~~
\begin{axopicture}(120,80)(0,15)
\SetWidth{1.5}
\Photon(0,60)(20,60){3}{3}
\Oval(60,60)(25,40)(0)
\Photon(100,60)(120,60){3}{3}
\Vertex(20,60){3}
\Vertex(100,60){3}
\PhotonArc(100,110)(50,210,270){3}{7.5}
\Vertex(53,85){3}
\end{axopicture}
\\
\begin{axopicture}(120,80)(0,15)
\SetWidth{1.5}
\Photon(0,60)(20,60){3}{3}
\Oval(60,60)(25,40)(0)
\Photon(100,60)(120,60){3}{3}
\Vertex(20,60){3}
\Vertex(100,60){3}
\PhotonArc(20,10)(50,30,90){3}{7.5}
\Vertex(67,35){3}
\end{axopicture}
~~
\begin{axopicture}(120,80)(0,15)
\SetWidth{1.5}
\Photon(0,60)(20,60){3}{3}
\Oval(60,60)(25,40)(0)
\Photon(100,60)(120,60){3}{3}
\Vertex(20,60){3}
\Vertex(100,60){3}
\PhotonArc(100,10)(50,90,150){3}{7.5}
\Vertex(53,35){3}
\end{axopicture}
~~
\begin{axopicture}(120,80)(0,15)
\SetWidth{1.5}
\Photon(0,60)(20,60){3}{3}
\Oval(60,60)(25,40)(0)
\Photon(100,60)(120,60){3}{3}
\Vertex(20,60){3}
\Vertex(100,60){3}
\PhotonArc(60,70)(15,0,360){3}{15}
\Vertex(60,85){3}
\end{axopicture}
~~
\begin{axopicture}(120,80)(0,15)
\SetWidth{1.5}
\Photon(0,60)(20,60){3}{3}
\Oval(60,60)(25,40)(0)
\Photon(100,60)(120,60){3}{3}
\Vertex(20,60){3}
\Vertex(100,60){3}
\PhotonArc(60,50)(15,0,360){3}{15}
\Vertex(60,35){3}
\end{axopicture}
~~
\begin{axopicture}(120,80)(0,15)
\SetWidth{1.5}
\Photon(0,60)(20,60){3}{3}
\Oval(60,60)(25,40)(0)
\Photon(100,60)(120,60){3}{3}
\Vertex(20,60){3}
\Vertex(100,60){3}
\Photon(20,60)(100,60){3}{12}
\end{axopicture}
~~
\begin{axopicture}(120,80)(0,15)
\SetWidth{1.5}
\Photon(0,60)(20,60){3}{3}
\Oval(60,60)(25,40)(0)
\Photon(100,60)(120,60){3}{3}
\Vertex(20,60){3}
\Vertex(100,60){3}
\Photon(60,35)(60,85){3}{8}
\Vertex(60,35){3}
\Vertex(60,85){3}
\end{axopicture}
\end{center}
\caption{\label{fig:diagramsHVP} The diagrams to two-loop order of the
vector two-point function in scalar QED. Lowest-order is the top line.
Not shown are the diagrams involving counterterms and the disconnected
contribution.}
\end{figure}

The two-point function can be calculated from the diagrams in Fig.~\ref{fig:diagramsHVP}
To be precise we calculated, with $t_{\mu\nu}$ the spatial part
of $g_{\mu\nu}$
\begin{align}
\widetilde\Pi((p^0)^2) \equiv
   \frac{-1}{3p^2}t_{\mu\nu}\left(\Pi^{\mu\nu}(p)-\Pi^{\mu\nu}(p=0)\right)
 \end{align}
 which reduces to $\Pi(p^2)$ in the infinite volume limit.
In terms of the functions
\begin{align}
\Omega_{ij}(p^2/m^2) =\int\frac{d^{d-1}l}{(2\pi)^{d-1}}\frac{m^{i+2j-d+1}}{({\vec l^2}+m^2)^{i/2}(4{\vec l}^2+4 m^2-p^2)^j}
\end{align}
we obtain
\begin{align}
\widetilde\Pi(p^2) =
+\frac{c_0}{m^3 L^3}
\Bigg(
          - \frac{16}{3}\Omega_{0,3}
          - \frac{5}{3}\Omega_{2,2}
          + \frac{40}{9}\Omega_{2,3}
          - \frac{3}{8}\Omega_{4,1}
          + \frac{7}{6}\Omega_{4,2}
          + \frac{8}{9}\Omega_{4,3}
\Bigg)
+\mathcal{O}\left(\frac{1}{L^4},e^{-mL}\right).
\end{align}
Note that it starts only at order $1/L^3$. This is because the two-point
function is a neutral object and far away the photon sees only the dipole
effect, not a charge. In the Euclidean all meson lines are off-shell as well.
We have checked that our analytical result agrees well with the same result
obtained using lattice perturbation theory for scalar QED as well as with a
numerical evaluation putting scalar QED on a lattice. Details can be found
in~\cite{Bijnens:2019ejw}.

\begin{figure}
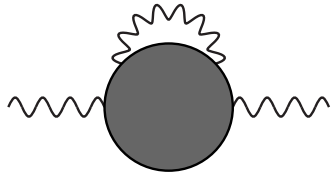

\begin{center}
\begin{axopicture}(100,80)(-10,0)
\SetScale{1.2}
\SetWidth{0.75}
\Photon(-10,20)(20,20){3}{3.5}
\Photon(60,20)(90,20){3}{3.5}
\PhotonArc(40,35)(15,-30,210){2.5}{9.5}
\GCirc(40,20){20}{0.4}
\end{axopicture}
\end{center}
\caption{\label{fig:HVPgeneric} A generic QED correction to the vector
two-point function.}
\end{figure}

The conclusion that the correction only starts at $1/L^3$ is general. The
correction can be seen as the generic diagram in Fig.~\ref{fig:HVPgeneric}.
If we cut open the photon line it reduces the blob to the vector four-point
function shown in Fig.~\ref{fig:hadronic}c. That four-point function has no
infra-red singularities, that can be seen from the analysis in
\cite{Colangelo:2017fiz} and references therein.
Gauge-invariance enforces at least a factor
cancelling the photon singular behaviour as discussed in more detail
in~\cite{Bijnens:2019ejw}. The $1/L^3$ behaviour is \emph{universal} though
the coefficient in front is not. For reasonable values of $L$ the finite
volume correction is thus expected to be small and negligible for the precision
needed in the near future.

\section{Conclusions}
\label{sec:conclusions}

This talk presented the main result of~\cite{Bijnens:2019ghy}
and~\cite{Bijnens:2019ejw}. Both concern hadronic contributions to the muon
anomalous magnetic moment. In the first we obtained that the usual quark-loop is indeed the first term in a systematic OPE. The next term was also calculated
and found to be numerically small. In the second we showed that the
electromagnetic finite volume corrections to the vector two-point function in
general only start at order $1/L^3$ and checked these results in scalar QED
with a continuum method, lattice perturbation theory and a numerical
lattice scalar QED calculation.

\section*{Acknowledgements}
This work is supported in part by the Swedish Research Council grants contract
numbers 2015-04089 and 2016-05996, and by the European Research Council
(ERC) under the European Union's Horizon 2020 research and innovation
programme under grant agreement No 668679.

\end{document}